\documentstyle[psfig]{mn}
\newif\ifAMStwofonts

\title[NGC~3801: AGN feedback caught in the act?]
      {NGC~3801 caught in the act: A post-merger starforming early-type galaxy with AGN-jet feedback}
\author[Ananda Hota et al.]
    {Ananda Hota$^1$,$\thanks{E-mail: hotaananda@gmail.com}$ 
Soo-Chang Rey$^2$, Yongbeom Kang$^{2,3}$, Suk Kim$^2$, 
\newauthor Satoki Matsushita$^1$, Jiwon Chung$^2$ \\
$1$ Academia Sinica Institute of Astronomy and Astrophysics, P.O. Box 23-141, Taipei 106, Taiwan, R.O.C. \\
$2$ Department of Astronomy and Space Science, Chungnam National University, Daejeon 305-764, Republic of Korea\\
$3$ Department of Physics and Astronomy, Johns Hopkins University, Baltimore, MD 21218, USA\\
}
\date{Accepted.    Received}
\pagerange{\pageref{firstpage}--\pageref{lastpage}}
\pubyear{  }
\begin{document}
\maketitle
\begin{abstract}
In the current models of galaxy formation and evolution, AGN feedback
is crucial to reproduce galaxy luminosity function, colour-magnitude relation
and M$_\bullet$$-\sigma$ relation. However, if AGN-feedback can indeed
expel and heat up significant amount of cool molecular gas and consequently 
quench star formation, is yet to be demonstrated observationally. 
Only in four cases so far (Cen~A, NGC~3801, NGC~6764 and Mrk~6), X-ray 
observations have found evidences of jet-driven shocks heating the ISM.
We chose the least-explored galaxy, NGC~3801, and present the first ultraviolet imaging and stellar population 
analysisis of this galaxy from GALEX data. We find this merger-remnant early-type galaxy to have an intriguing 
spiral-wisp of young star forming regions (age ranging from 100--500 Myr). 
Taking clues from dust/PAH, H{\sc i} and CO emission images we interpret NGC~3801 
to have a kinamatically decoupled core or an extremely warped gas disk. 
From the HST data we also show evidence of ionised gas outflow similar to that observed in H{\sc i} and 
molecular gas (CO) data, which may have caused the decline of star formation leading to the red optical colour of the galaxy. 
However, from these panchromatic data we interpret that the expanding shock 
shells from the young ($\sim$2.4 million years) radio jets are yet to reach the 
outer gaseous regions of the galaxy. It seems, we observe this galaxy at a rare stage of 
its evolutionary sequence where post-merger star formation has already declined and new powerful 
jet feedback is about to affect the gaseous star forming outer disk within the next 
10 Myr, to further transform it into a red-and-dead early-type galaxy.        
\end{abstract}
\begin{keywords} 
galaxies: active -- galaxies: evolution -- galaxies: individual: NGC~3801
\end{keywords}
\section{Introduction}
By incorporating feedback from star formation and AGNs into cosmological simulations, 
various puzzling statistical correlations of galaxies like the galaxy luminosity function, the
$M_\bullet$$-\sigma$ relation, and the colour-magnitude bi-modality of nearby
galaxies can be reproduced (Springel, Di Matteo \& Hernquist 2005, Croton et al.
2006). In the models of galaxy evolution via merger, the abrupt separation
between spiral, gas-rich, young star forming galaxies and their merger product
gas-poor, old stellar population, elliptical galaxies requires AGN feedback to be
playing a decisive role (Springel, Di Matteo \& Hernquist 2005). Statistical 
studies of early-type galaxies (ETG) from Sloan Digitised Sky Survey (SDSS) and 
Galaxy Evolution Explorer (GALEX) data argue in favour of a fast (within a billion year)
quenching of star formation process driven by the Active Galactic Nuclei (AGN) feedback 
(Schawinski et al. 2007,
Kaviraj et al. 2011). However, a direct observational evidence for an AGN-driven gas outflow
or heating to have quenched the star formation process in any galaxy is still missing. 

If in a sample of post-merger ETGs, star formation is found to be on
the decline, mass outflow rate is larger than the ongoing star formation rate
and AGN feedback timescale coincide with any sharp break in the star formation
history, then the feedback-driven galaxy evolution hypothesis will get a substantial
support. The UV-bright ETGs and the post-starburst
galaxies could be the prime targets to look for relics of past AGN feedbacks.
In this case, radio loud AGNs are probably the only kind of AGNs where 
the relics of past AGN feedback, as relic lobe, can be investigated.     
Radio jets, often episodic in a few million years (Myr) timescale, may drive frequent
and powerful gaseous outflows and shock-heating of the Interstellar Medium (ISM) 
to prevent star formation. From Chandra X-ray observations, so far, only four
cases are known to show galactic-scale shocks driven by AGN-produced radio lobes, 
which are Cen~A, NGC~3801, NGC~6764 and  Mrk~6 (Mingo et al. 2011 and references therein). 
We chose to analyse the least explored galaxy, NGC~3801 (Croston et al. 2007), 
which is also the clearest case of kpc-scale shocks 
apparently buried within the rich ISM of the host galaxy, in an attempt to investigate 
galaxy evolution via merger, star formation and AGN-feedback processes.
\begin{figure}
\vbox{
  \psfig{file=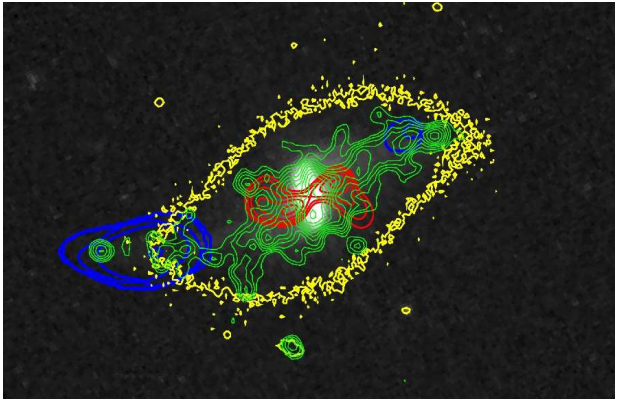,width=3.2in,angle=0}
  }
\caption[]{Panchromatic summary of NGC~3801 showing radio continuum (red), H{\sc i} emission (blue), 
8 $\mu$m dust/PAH emission (gray scale), SDSS r$^{\prime}$ band single contour (yellow) 
and near UV (green) images, from various observations (Hota et al. 2009). 
See Das et al. 2005 and Croston et al. 2007 for CO and X-ray emission images, respectively.  
              }
\end{figure}

{\bf NGC~3801:} It has been known since very early time that it is an FR I, low-power, 
compact radio source ($\sim$40$^{\prime\prime}$ or $\sim$9 kpc) seen smaller than 
the optical host galaxy and likely to be buried within the ISM. The host galaxy has been 
classified S0/a and morphology is disturbed. It has one minor-axis dust-lane and 
another prominent dust filament extending from the centre to the eastern edge of 
the galaxy, roughly orthogonal to the dust-lane (Heckman et al. 1986, and references therein). 
The faint light distribution on the outer edges was also known to show boxy
shape and a stellar tail on the south-eastern corner of the box. 
Within the box shape, the stellar light show an open S-shaped structure, more 
similar to the hysteresis loop. The radio lobes, also in the shape of
S, is seen confined within this hysteresis loop structure (HLS), with similar
clockwise bending (See Fig.1; Heckman et al. 1986).  

Surrounding these radio lobes, shock shells have been discovered with the Chandra X-ray 
observations (Croston et al. 2007). These shock shells have been estimated to be expanding at a velocity of 
850 km s$^{-1}$ and material driven away by it may escape from the ISM of the galaxy. 
Cool molecular gas
emission, traced by CO (1-0), has been imaged by Das et al. (2005). This revealed three clumps,
two roughly on the minor-axis dust lane, suggesting rotating gas disk (r = 2 kpc), and the third 
clump on the eastern filament at similar radial distance. Atomic hydrogen emission 
(H{\sc i} 21cm line), first detected by Heckman et al. (1983), has recently been imaged by 
Hota et al. 2009 (hereafter Hota-09), who suggest a 30 kpc size gas disk rotating along the major axis 
of the galaxy (Fig. 1, blue contours). 

Ultraviolet (UV) light traces the young star formation
and has potential to tell stories of the last major episode of star formation in an 
ETG and its interplay with the AGN. In this {\it letter} we present the first UV imaging study 
and investigation of the stellar population of this galaxy, from the archival GALEX data. 
We discuss our findings of this interesting target, a prime
candidate to understand galaxy evolution by merger, star formation and AGN feedback,
incorporating all available panchromatic information. For a systemic velocity of 
3451 km s$^{-1}$, we adopt a distance of 47.9 Mpc to NGC~3801, corresponding to the
angular scale of  1$^{\prime\prime}$ = 0.23 kpc. For easy comparison with X-ray and molecular 
gas images published earlier we have presented all the images in B1950 co-ordinate system. 
\section{UV observations and data analysis}
NGC~3801 was imaged with GALEX in both the far-ultraviolet (FUV;
1350--1750 \AA) and near-ultraviolet (NUV; 1750--2750 \AA). We extracted the UV images
of NGC~3801 from the GALEX Release 5 (GR5) archive. The total integration times
were 111 s and 265 s for FUV and NUV, respectively. The GALEX
instruments and their on-orbit performances and calibrations are described by 
Morrissey et al. (2007 and references therein).
GALEX FUV and NUV imaging has 4$^{\prime\prime}$.2 and
5$^{\prime\prime}$.3 resolution (FWHM) which corresponds to $\sim$ 0.97 kpc and 1.2 kpc, respectively.
In NUV and FUV images, we defined UV bright clumps to trace their morphology
and to measure their UV flux densities. The images were adaptively smoothed
using the Gaussian filter by 3 pixel ( $\sim$5 $^{\prime\prime}$)
radius in order to select morphologically
distinct clumps. Various thresholds above the mean sky background value
were applied for delineation of the clumps. We detect 9 clumps in the smoothed NUV
image each of which has at least 10 contiguous pixels. We measured UV flux
within the contours of the clumps, and subtracted the local background using the
GALEX GR5 pipeline background image (``-skybg").
Finally, FUV and NUV magnitudes in the AB magnitude system and photometric errors
of all clumps were determined (see Table 1). Foreground Galactic extinction was
corrected using the reddening maps of Schlegel, Finkbeiner \& Davis (1998) and
the extinction law of Cardelli, Clayton \& Mathis (1989).
\section{Results}
\begin{figure}
\vbox{
  \psfig{file=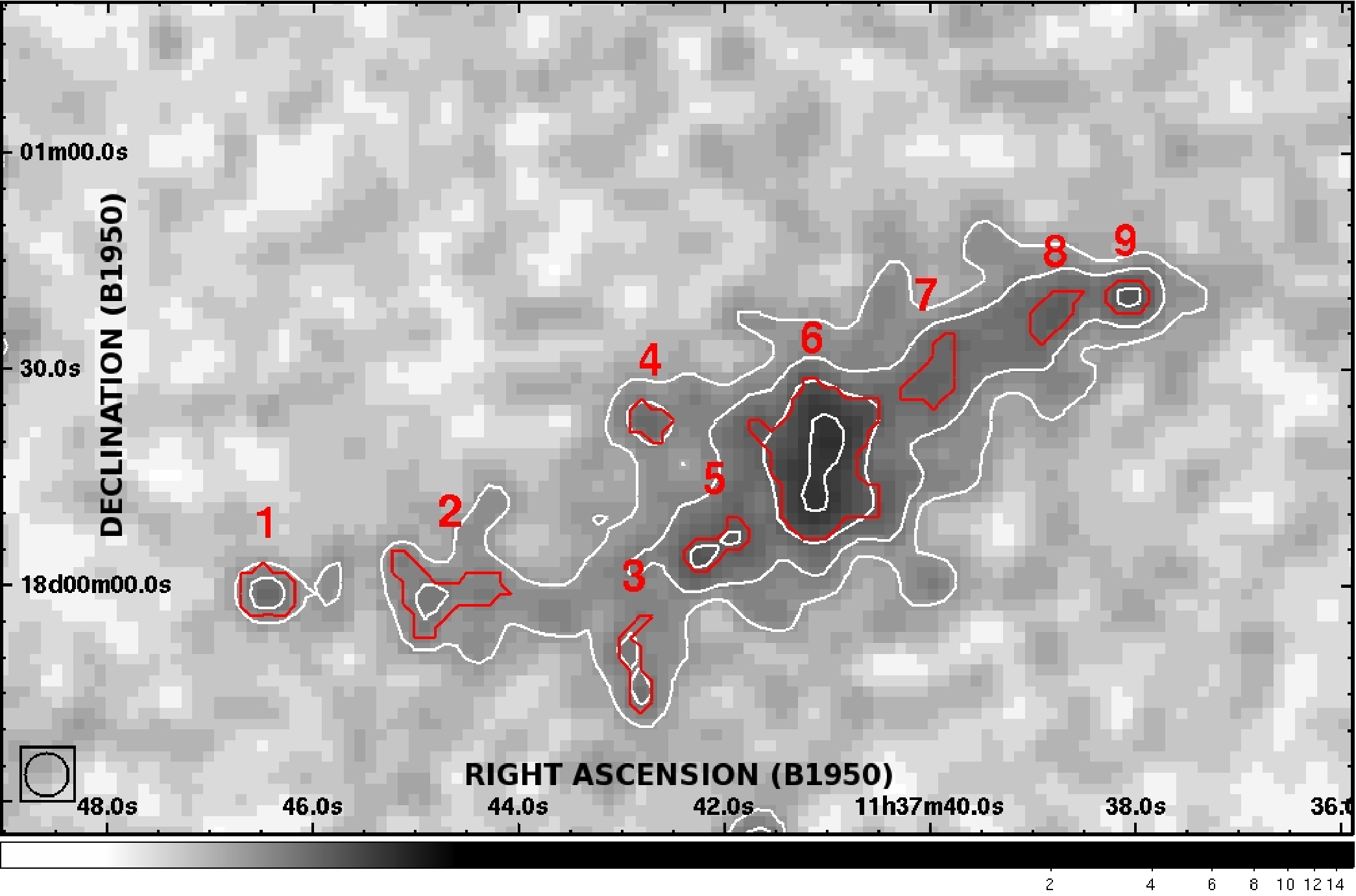,width=3.0in,angle=0}
   \psfig{file=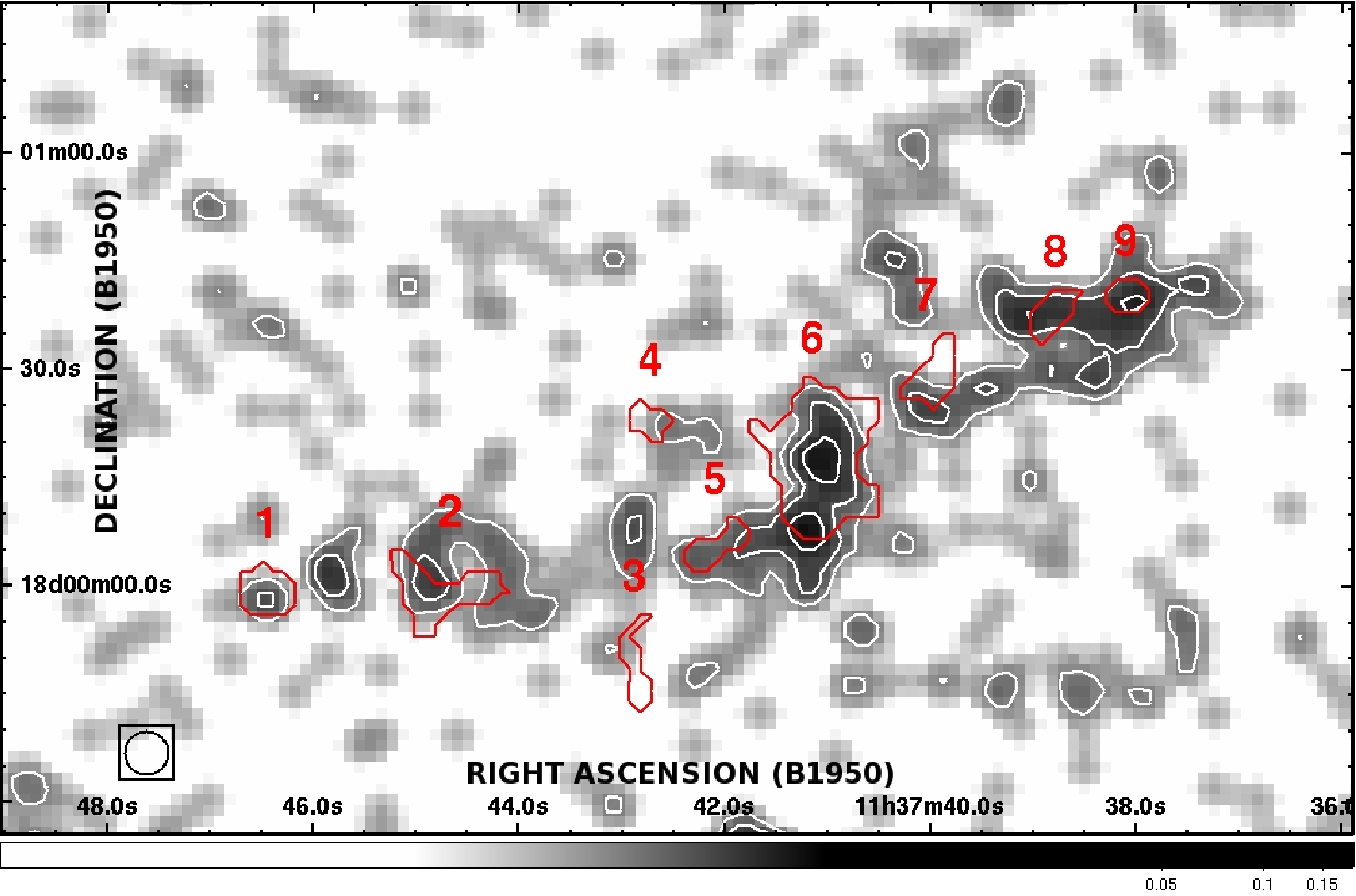,width=3.0in,angle=0}
  }
\caption[]{
Smoothed near UV (top panel) and far UV (bottom panel) image of NGC~3801, presented both in linear gray scale and white contours. 
Regions selected for stellar population age-analysis, from the NUV image, have been marked in red contours and labelled 1 to 9.
              }
\end{figure}
 \begin{table}
 \caption{Photometry and Age Estimation of the UV Clumps}
 \hspace{-0.1cm}  \begin{tabular}{@{}ccccc@{}}
\hline
clump&  far UV            &  near UV        &Age Myr           &Age Myr  \\
no.&    AB mag            &  AB mag         &Z=0.02            &Z=0.05            \\
\hline                                                                          
 1 &      22.678$\pm$0.433& 21.808$\pm$0.175&  371$\pm$131     &332$\pm$134     \\
 2 &      21.921$\pm$0.304& 21.135$\pm$0.129&  349$\pm$90      &311$\pm$93      \\
 3 &      23.555$\pm$0.699& 21.842$\pm$0.175&  527$\pm$127     &475$\pm$100     \\
 4 &      23.430$\pm$0.620& 22.164$\pm$0.200&  455$\pm$142     &416$\pm$128     \\
 5 &      22.122$\pm$0.320& 21.478$\pm$0.142&  311$\pm$131     &289$\pm$140     \\
 6 &      20.419$\pm$0.148& 19.364$\pm$0.053&  413$\pm$33      &390$\pm$32      \\
 7 &      23.045$\pm$0.532& 21.510$\pm$0.148&  499$\pm$97      &458$\pm$84      \\
 8 &      21.929$\pm$0.290& 21.756$\pm$0.164&  111$\pm$469     & 85$\pm$364     \\
 9 &      22.094$\pm$0.311& 21.915$\pm$0.174&  114$\pm$509     & 92$\pm$409     \\
\hline
10 &      18.724$\pm$0.071& 17.696$\pm$0.026&  408$\pm$16      &362$\pm$15      \\
\hline
\end{tabular}\hfill\break
\end{table}
{\bf UV morphology:} The smoothed NUV image shows an intriguing
S-shaped morphology (Fig. 2 top panel). In contrast to the optical morphology 
(Fig.1 yellow contour), where the envelope of the stellar light was HLS, here this S-shaped
NUV spiral-wisp is seen as the central spine of the optical HLS.
Some diffuse and fainter emission is also seen in the intermediate region
between spiral-wisp and the HLS. The western wisp is brighter than
its eastern counterpart, but they extend up to similar distances
(60$^{\prime\prime}$--70$^{\prime\prime}$ or 14--16 kpc) from the centre of the galaxy. NUV wisps
connect to an elongated bright structure at the centre, oriented nearly north-
south. It is as bright in surface brightness as any compact emission on the spiral-wisp.
This central elongated structure does not show any strong peak at the centre,
but extends $\sim$15$^{\prime\prime}$ (3.5 kpc) on either side of the centre. This elongated
emission can be compared with the minor axis dust-lane but its detail
structure and orientation differs. Although the major axis
extension in optical and UV are comparable, the UV emission along the
minor-axis is smaller, nearly half of the optical HLS (Fig. 1).
We also present the FUV image smoothed to same resolution
(Fig. 2 bottom panel). Although this shows similar spiral-wisp
structure as in NUV, it looks clumpier and noisier. Unlike in NUV,
the FUV emission is centrally peaked 
(R.A. 11h 37m 41.1s Dec. +18$^\circ$ 00$^{\prime}$ 16$^{\prime\prime}$),
and is closer to the centre of the galaxy 
(R.A. 11h 37m 41s.2 Dec. +18$^\circ$ 00$^{\prime}$ 18$^{\prime\prime}$; 
derived from IR images). Since UV traces the recent massive star
formation, this spiral-wisp, seen as a curved spine to the HLS,
is likely a distinct newly formed structure containing relatively young stars.

{\bf Age of the stellar population:} Small clumps along the spiral-wisp
labelled ‘1’ to ‘9’ (Fig. 2) and the whole emission region, for the galaxy,
assigned a number ‘10’ have been chosen for extracting photometric parameters.
It is unclear if clumps ‘1’ and ‘9’ are part of NGC~3801 or
independent galaxies, therefore for our analysis they have been excluded from
determining global parameters of NGC~3801. Although there is no complete
overlap between the NUV and FUV images for a few clumps, still we went ahead in our analysis
for a representative value of the age of stellar population.
We estimated luminosity-weighted average ages for the stellar population
in clumps along the spiral-wisp by comparing FUV-NUV color with synthetic
population models of Bruzual \& Charlot (2003). In Table 1 we present the ages 
of the clumps derived based on the models of simple stellar population (i.e. single
burst star formation). Two different metallicity values, Z = 0.02 (Fe/H = +0.0932) 
and Z = 0.05 (Fe/H = +0.5595) were assumed for the age estimations and listed in separate columns. 
The errors are calculated using propagation of photometric errors. 
The stellar population age, ranges from 100 to 500 Myr for metallicity Z = 0.02
and slightly lower for a higher metallicity.
Clump 8 associated with bright UV emission region on the western spiral-wisp,
shows very young, 85--110 Myr stellar population, although the error bars
are very large. Clumps 2, 3, 5, 7, and 8 show an average age $\sim$300 Myr,
demonstrating recent star formation along the spiral-wisp.
No significant trend of stellar age with the distance from the nucleus was
noticed. 
The central 20$^{\prime\prime}$ (4.6 kpc) region of the galaxy (clump 6) with typical
age of 400 $\pm$ 30 Myr, ignoring the effect of the AGN represent the most
luminous star forming region. Since NGC~3801 as a whole has a similar
$\sim$ 400 Myr age stars, this suggests a recent star formation in this early-type
radio galaxy host with many peculiar optical morphologies.
Two clumps ‘1’ and ‘9’, as described earlier, show average stellar age of
$\sim$ 350 $\pm$ 130 Myr and 100 $\pm$ 450 Myr respectively. Coincidental location
at the ends of the spiral-wisp and similarity of stellar ages of
these two clumps with that of the spiral-wisp, suggests that they are
likely young star forming regions associated with NGC~3801.
\section{\bf DISCUSSION}
\begin{figure}
\hbox{
  \psfig{file=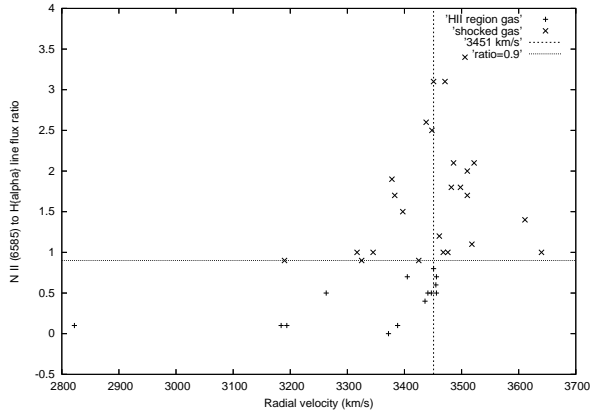,width=3.1in,angle=-90}
  }
\caption[]{N{\sc ii} (6585) to H$\alpha$ line flux ratios have been plotted
against the radial velocity of the emission components. Data has been taken from
Noel-Storr et al. (2003). The lower ($<$0.9) and higher line ratios are marked 
with `+' and `x' signs, respectively. The horizontal and vertical lines mark
ratio = 0.9 and V$_{\rm sys}$ = 3451 km s$^{-1}$, respectively.    
              }
\end{figure}
{\bf Origin of the spiral-wisp:} 
To understand the origin of the young star forming UV bright spiral-wisp, we need
to take clues from the imaging and spectroscopy done at other wavelengths.
Hence, in Fig. 1, we have summarised the 
NUV emission map, from the GALEX (green contours), 
optical r$^{\prime}$ band single contour image of the HLS, from the SDSS (yellow contours), 
dust and PAH emission image at 8 $\mu$m, from the {\it Spitzer Space Telescope} (gray scale), 
and H{\sc i} emission moment-0 map (blue contours) and 1.4 GHz radio continuum image, 
both from the Very Large Array (red contours),
of the galaxy for a panchromatic analysis. Readers are referred to Das
et al. (2005) and Croston at el. (2007) for CO emission and X-ray emission
images, respectively. 
The spiral-wisp clearly fits in as central spine to the HLS. 
The HLS, an atypical stellar distribution in an ETG like NGC~3801, 
deserves a kinematic modeling to be explained properly. 
On the eastern and western ends of the HLS, blobs
of rotating H{\sc i} represent a $\sim$30 kpc size gas disk (Hota-09). On the other
hand, within the central 4 kpc, CO emission clumps suggest rotating molecular
gas disk oriented roughly north-south (Das et al. 2005). Two orthogonally rotating 
gas disks can not co-exist, radial location has to be different. 
This resembles a gas dynamical condition as seen in 
Kinematically Distinct/Decoupled Cores (KDC; Barnes 2002). 
In this KDC interpretation, the  red-shifted eastern molecular cloud clump of Das et al. 
could be re-interpreted as part of the outer gas disk rotating along the optical major axis.
As mentioned in Hota-09, although the H{\sc i} rotates in the same sense as the
stars, gas rotate twice faster than the stars. Both these forms of decouplings reinforce the
idea of a merger origin of NGC~3801 (Barnes 2002, Heckman et al. 1986). 

Alternate to a KDC, it can also be interpreted as an
extremely ($\sim$90$^{\circ}$) warped gas disk. 
The central north-south gas disk changes its orientation to east-west, 
with the increase of radius from 2 kpc to 15 kpc. 
The central 2 kpc molecular gas disk has a corresponding bright
linear feature in the UV as well as in the dust/PAH emission. Similarly, the 
full spiral-wisp of UV emission also has a correspondence in the dust/PAH emission 
observed by the {\it Spitzer} (Fig. 1, Hota-09, Hota et al. in preparation). 
The HLS may possibly be reproduced if light from a low level star formation 
in the warped gas disk is added onto the dominant elliptical distribution of 
stellar light. Furthermore, if the gas disk is twisted or due to the projection effect, 
an inclined warped disk can re-produce the spiral-wisp of higher density and younger stars
just along the line of orbital crowding. 
The brighter western part of the spiral-wisp is
likely in the foreground and the eastern counterpart in the background. 
A direct comparison of the spiral-wisp can be seen in the
galaxy NGC~3718, where the warped, inclined and twisted gas disk has been kinematically
modeled in detail using H{\sc i} emission velocity field (Sparke et al. 2009). More sensitive 
H{\sc i} and CO imaging of NGC~3801 will be very helpful to test our proposal.

{\bf State of star formation:}
As the last major phase of star formation in the galaxy happened $\sim$400 Myr ago, 
the star formation process now is on the decline. When UV satellite data is not available, 
u$^{\prime}$-r$^{\prime}$ optical colour of galaxies, is often used as a proxy for starformation 
history, in large sample studies (e.g. Schawinski et al. 2007). 
The u$^{\prime}$-r$^{\prime}$ colour and r$^{\prime}$ band absolute magnitude of NGC~3801, as listed in the DR7, are 3.08 
and $-$21.27. This puts NGC~3801 clearly in the location of 
`red sequence' or `red-and-dead' type of galaxies (Strateva et al. 2001, Bell et al. 2004). 
From a large sample study of morphologically selected ETGs from the SDSS, 
Schawinski et al. (2007) proposed an evolutionary sequence in which star forming 
ETGs quickly transform to a quiescent red-galaxy phase via. an increasingly 
intense AGN activity over a period of one billion year. 
Young star formation in ETGs are usually found 
in the form of UV-bright rings and the spiral-wisp seen in NGC~3801 is bit unusual (Salim \& Rich 2010). 
An important question arises, if this unusual UV morphology as well as the decline of star 
formation activity is related to any possible feedback processes from a past AGN or 
starburst superwind, and the galaxy is caught while changing. 

{\bf Signatures of feedback:} 
The eastern molecular gas clump seen superposed on the eastern jet was 
interpreted as partially jet-entrained (outflow) gas by Das et al. (2005).
Presence of broad blue-shifted H{\sc i}-absorption line seen against the 
same eastern radio jet was also mentioned by Hota-09 as a sign of outflow. 
We looked for further kinematic signatures of gaseous feedback in it. 
In Fig. 3 we plot the radial velocity of the emission line components and their
N{\sc ii} (6585) to H$\alpha$ line flux ratio, extracted from the central
1$^{\prime\prime}$.8 $\times$0$^{\prime\prime}$.4 (0.4$\times$0.1 kpc)
region of NGC~3801 
(from the {\it Hubble Space Telescope} (HST) data in the table 14 of Noel-Storr et al. 2003).
Components with line ratio less than 0.9, which would suggest H{\sc ii}-region
like ionised gas property, show a different distribution from components with
higher line ratio representing ionisation by shocks (e.g. Veilleux \& Rupke
2002).  While the higher line ratio shocked gas is symmetrically
distributed around the $V_{sys}$, the lower line-ratio gas is either near
$V_{sys}$ or significantly blue-shifted. The maximally blue-shifted 
(by 630 km s$^{-1}$) component also has the highest H$\alpha$ line flux (21.4 $\times$
10$^{-16}$ erg s$^{-1}$ cm$^{-2}$ Hz$^{-1}$) and the highest velocity
dispersion (205 km s$^{-1}$). Absence of red-shifted component is likely
due to extinction but the highly blue-shifted components clearly suggest 
outflow of ionised gas from the very central region. Whether this is directly related to 
AGN wind/jet, starburst superwind or outwardly moving shells during merger activity 
is unclear.

{\bf State of AGN activity:}  Schawinski et al. (2007) suggest that AGN activity peaks
$\sim$500 Myr after the peak star formation activity. With a glance through the SDSS DR7 spectral 
line data, we found that NGC~3801 can be classified in between Seyferts and LINERs 
(log(N{\sc II}/H$\alpha$) = 0.138 and log(O{\sc III}/H$\beta$) = 0.538.). This particular fact, 
u-r colour and the stellar velocity dispersion of  225 km s$^{-1}$ (de Nella et al. 1995), 
suggest that NGC~3801 is near the last phase of AGN activity prior to reaching a quiescent
phase as per the evolutionary plot of figure 7 in  Schawinski et al. (2007). 
There is no clue as to how long the AGN activity has been in existence. Spectral ageing of the 
radio lobes is probably the only method of age-dating AGNs, if radio-loud. Croston et al. have 
estimated the spectral age from the steepest regions in the extended lobes 
or wings, where the oldest radio plasma is expected. Their spectral age of $\sim$2.4 Myr
is only suggesting a time-scale since when fresh relativistic plasma supply to the wings has been
stopped. Hence, it is possible that wings may have been in existence since several tens of Myr. 
We also do not know the real orientation of the wings of the radio lobes, it may be much longer than 
what we see in projection. On the other hand the straight part of the radio jets, are visible in
the sub-mm wavelengths, from which Das et al. interpret that the jets are expanding in the
plane of the sky.   

In the figure 2 of Croston et al. (2007) it is clear that the wings extends farther than the 
X-ray emitting shock shells. If wings are deflected hydrodynamic flow of the jets, they can not travel
farther than the shock front. Hence, wings are older and shock is associated only with the 
current (younger than 2.4 Myr) jet activity. 
This jet is clearly orthogonal to the central
minor-axis gas disk (r = 2 kpc). In a case (3C~449) where nuclear dust disk is warped, 
Tremblay et al. (2006) has discussed about the cause-effect ambiguity between the 
jet to have shaped the warped dust/gas disk or the reverse. 
Gopal-Krishna et al. (2010) suggest rotating ISM as a possible cause of wing formation, 
taking NGC~3801 as an example. As proposed for X-shaped radio galaxies (Gopal-Krishna et al. 2010), 
it is also possible that a recent binary black hole coalescence may have caused fast 
(within last 2.4 Myr) change of jet axis in NGC~3801.   

{\bf Caught in the act:} 
It is intriguing that the eastern dust filament, stops at the centre, with no 
extension to the west. Symmetry is expected due to rotation along the major axis.
The H{\sc i}-gas rotation period of $\sim$330 Myr (for r = 15 kpc and v = 280 km s$^{-1}$; Hota-09)
suggest that these dust filament and star forming UV-clumps are likely too young to have east-west symmetry.  
Similarly, lack of UV emission on the north-eastern and south-western quadrant of the galaxy, 
where wings of the radio lobes are seen, on projection, also present a curious co-incidence.
In the fine dust silhouette image from the HST data (figure 5 of Das et al.) there is 
no signature of the effect from the radio jet, wing or shock shells. As the shock is 
associated only with the jet (younger than 2.4 Myr) expanding in the plane of the sky,
it is very likely that it is yet to reach the outer parts of the host galaxy where dust filaments
are present. In the warped gas disk geometry, the jets and shock are 
also likely to miss impacting on the outer gas disk in the early-phase. However, with 
an expansion velocity of 850 km s$^{-1}$ the shock can move 8 kpc 
in the next 10 Myr and can affect the outer gas (H{\sc i} and CO) and dust distribution. 
It is unclear, if the past AGN activity had an effect on the host galaxy gas distribution 
and star formation; however, the 
current jet-induced shock shells are powerful enough to do it. We are likely witnessing 
this merger-remnant ETG just 10 Myr prior to its AGN-jet feedback start 
disrupting its outer gas disk and potentially quench its star formation.  

{\bf Future directions:} 
Since the star formation has started declining a few 100 Myr back and the oldest 
plasma in the radio lobes is only 2.4 Myr old, we fail to connect if AGN-feedback is 
the cause behind the declining/quenching of star formation. 
Future studies need to include objects where AGN feedback and star
formation processes are on comparable time-scale and directly interacting to 
investigate the causal connection. 
For example, in the radio bright Seyfert galaxy NGC~6764, the central 2 kpc radio bubble outflow 
and its associated shock-heated hot gas, warm ionised gas, cool molecular and H{\sc i} gas outflow
has a dynamical timescale of 10--20 Myr (same as the precession timescale of the radio jet).
Interestingly, this is coincident with a break in the star formation history of the central region, 
which occured 5--15 Myr ago. It is tempting to ask if the outflow has caused the break in star formation 
(Hota \& Saikia 2006, Croston et al. 2008, Kharb et al. 2010). 
There is a need of observational programmes to bridge the gap by catching the recent most star
formation history with optical spectroscopy, molecular gas loss history say with ALMA,
and hunt for the radio relics of the oldest (up to several hundred Myr) episodes of 
AGN feedback. 
Deep low frequency radio surveys, like the ongoing 
TGSS
at 150 MHz and upcoming 
LOFAR sky surveys, may reveal relics of past AGN feedbacks in many post-starburst 
(Goto T. 2005), 
 young star forming ETGs to better understand merger and feedback 
driven galaxy evolution (see a potential demonstrator in Hota et al. 2011). 

\end{document}